# From Observation to Intervention: Memory in Brains and Large Language Models

Morteza Salehjahromi[1], Shayan A. Zadegan[2], Amgad Muneer[1], Jia Wu[1,3,4]✉

1. Department of Imaging Physics, The University of Texas MD Anderson Cancer Center, Houston, TX, USA
2. Department of Neurology, The University of Tennessee Health Science Center, Memphis, TN, USA
3. Department of Thoracic/Head and Neck Medical Oncology, The University of Texas MD Anderson Cancer Center, Houston, TX, USA
4. Institute for Data Science in Oncology, The University of Texas MD Anderson Cancer Center, Houston, TX, USA

## Abstract

Brains and large language models (LLMs) are fundamentally different memory systems, but they can be compared through shared functional questions: where memory-related information is represented, how partial cues recover broader associations, how new information is written or updated, and how memory-related states can be perturbed. In biological systems, these questions span synapses, neuronal ensembles, hippocampal–cortical interactions, and plasticity; in LLMs, they span weights, activations, context windows, retrieval systems, and external stores. The comparison is therefore functional and experimental rather than anatomical. Human studies reveal sparse concept responses, temporal binding, rapid association formation, episode-specific coding, and recall-related reactivation, but selective intervention remains limited. Rodent studies provide more selective causal access to learning-related ensembles, whereas human and macaque interventions usually affect broader circuits. LLMs lack lived episodic memory, yet they permit unusually direct and repeatable manipulation of internal states and stored information. We argue that this asymmetry creates a new opportunity. LLMs are not ahead in memory itself, but in experimental access. Their tools may help turn broad questions about retrieval, updating, persistence, reversibility, and unintended effects into sharper biological hypotheses. The productive bridge is to transfer experimental logic, not anatomical parts.

## 1. Introduction: compare the questions, not the anatomy

Brains form memories from embodied events and integrate perception, emotion, context, action, and personal history. LLMs learn statistical structure from data and generate outputs using parametric knowledge, temporary context, and, increasingly, external retrieval systems. A model component should therefore not be called an artificial hippocampus merely because it participates in retrieval. The goal is not a one-to-one comparison of anatomy, but a comparison of experimental questions: where information is represented, how partial cues retrieve broader associations, how new information is written or updated, and what changes when memory-related states are perturbed.

The evidence is markedly asymmetric. Human single-neuron studies reveal invariant concept responses, temporal binding, rapid association formation, episode-specific coding, and reactivation before verbal recall.[1-5] Yet these observations are rarely followed by selective manipulation of the same neurons. Rodent engram studies can tag, activate, silence, or reassociate neuronal ensembles recruited during learning.[6-8] Macaque and human experiments usually perturb broader hippocampal or entorhinal circuits, and their effects depend strongly on target, timing, frequency, and task.[9-11] LLM studies, by contrast, can patch activations, edit weights, steer representation directions, and rewrite external stores while measuring consequences throughout the system.[12-16]

This greater access does not mean that LLMs possess better memory. The edited object is usually a factual association, transient computational state, behavioral tendency, or stored record, not a lived episode embedded in perception and emotion.[17] The comparison becomes scientifically useful only when that distinction is explicit. This perspective therefore uses brains and LLMs as different systems that expose related questions at different levels of experimental control. We argue that controllable interventions in LLMs may help formulate sharper, testable hypotheses about biological memory, including retrieval, updating, persistence, and reversibility. **Figure 1** summarizes this asymmetry in experimental access, while **Box 1** defines the key terms used throughout the paper.

**Box 1**: Key memory terms used in this Perspective.

These definitions do not imply equivalent mechanisms in biological and artificial systems.

| Term | Meaning in this Perspective |
|---|---|
| **Lived episodic memory** | Memory of personally experienced events and their sensory, spatial, temporal, and emotional context. LLMs do not possess this form of memory. (**Biological**) |
| **Engram or learning-related ensemble** | A group of neurons recruited or changed during learning whose later activity contributes to memory expression. (**Biological**) |
| **Memory reactivation** | The reappearance of a memory-related neural pattern during rest or retrieval. (**Biological**) |
| **Parametric knowledge** | Information learned during training or fine-tuning and reflected in model weights. (**LLM**) |
| **Context-window information** | Information temporarily available in the current prompt or conversation; it guides generation without changing model weights. (**LLM**) |
| **External memory** | A separate, editable store, such as documents, databases, or knowledge graphs, accessed without changing model weights. (**LLM system**) |
| **Activation state** | A temporary internal pattern produced during inference that normally disappears afterward. (**LLM**) |

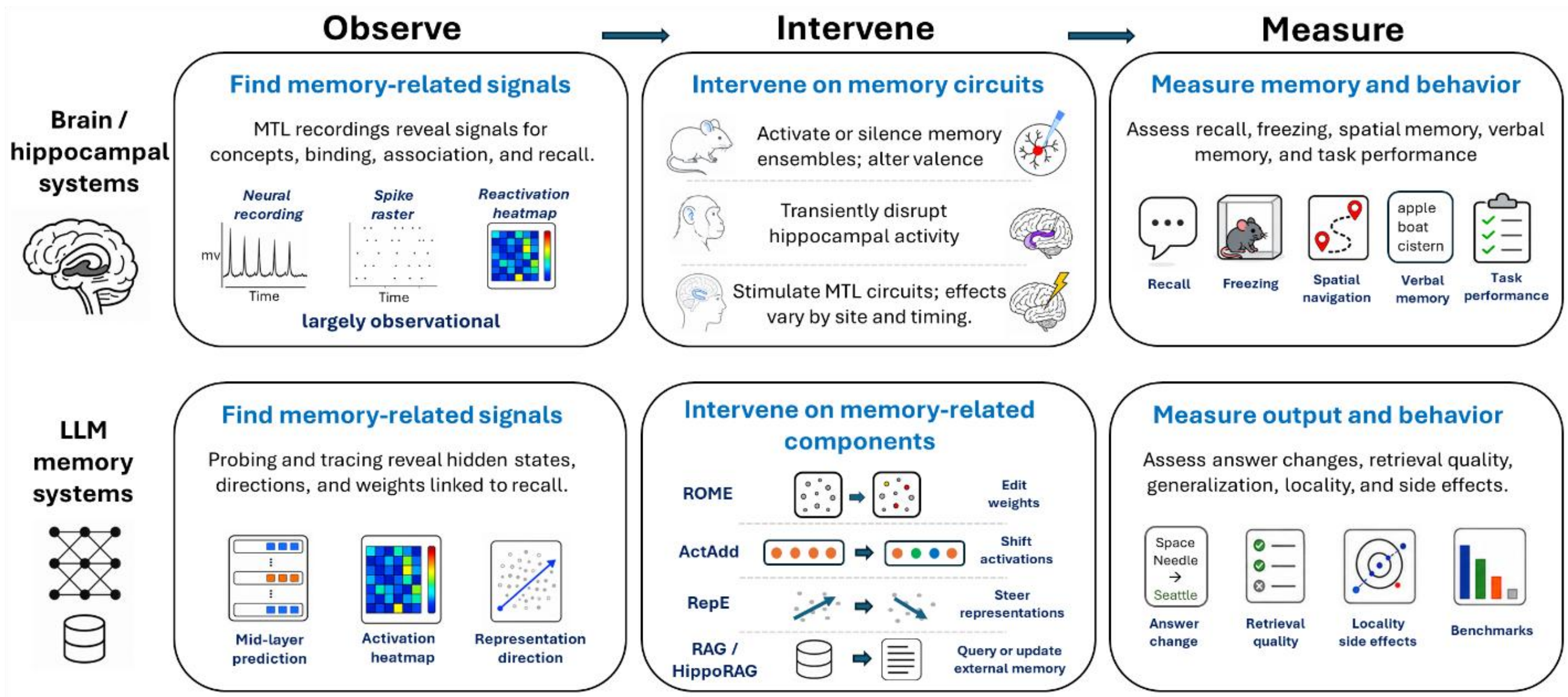

**Figure 1**: From observation to intervention in brain and LLM memory research. Human and macaque studies provide rich phenomenology but mostly broad intervention; rodent studies offer the strongest biological selectivity; and LLMs offer the most direct access to artificial weights, activations, representation directions, and external stores.

## 2. From observation to intervention: what changed in the story

The story begins with theories about how memories are organized and preserved. Hippocampal indexing theory proposed that the hippocampus can link information represented across cortex, allowing a partial cue to recover a broader memory.[18] Complementary learning systems then separated rapid acquisition from slower integration, framing memory as a balance between plasticity and stability.[19] Related design principles appear in artificial systems that separate relatively stable parametric knowledge from temporary context or external retrieval.[14,20] In both cases, new information must become accessible without erasing previously acquired knowledge.

Human and macaque recordings gradually showed how memory-related information appears in the activity of individual neurons. Macaque hippocampal neurons changed their responses as the animals learned new scene–location associations.[21] Human medial temporal-lobe recordings then revealed neurons that responded selectively to particular concepts across different images,[3] became active again before recall,[2] linked events that occurred close together in time,[4] rapidly learned person–place associations,[5] and responded to specific combinations of elements within an episode.[1] Collectively, these studies moved the field from asking whether a brain region is involved in memory to asking what specific information is represented by individual neurons and groups of neurons. However, seeing a memory-related signal does not show exactly what it does. A neuron or activity pattern may help store information, retrieve it, signal familiarity or confidence, or prepare a response. Its role may also change over time. Therefore, researchers need interventions to test whether the signal is required for memory, can produce a memory effect, or is only associated with successful recall.

Causal experiments developed differently across species. In rodents, researchers can label and selectively activate, silence, or change groups of neurons involved in learning.[6-8] In macaques and humans, drugs or electrical stimulation usually affect broader brain circuits. These studies showed that stimulation within hippocampal–entorhinal circuits can either improve or impair memory. The outcome depends not only on the location, but also on when and how the intervention is applied, the task being performed, the brain's current state, and how many neurons are affected.[9-11] These systems should not be ranked as if one is always better. Rodent studies can target learning-related neuron groups more precisely. Macaque and human studies, in contrast, can examine more complex behavior, and human studies can also use verbal reports, although their interventions are usually broader. Each system therefore answers a different part of the question. In general, greater intervention precision often comes with less behavioral complexity.

More recently, LLM research provided a different kind of experimental access. Researchers showed that feed-forward layers can act like key–value memories,[12] factual associations can be located and edited,[15] activation states can be steered,[13,16] and graph-based retrieval systems can be built using ideas from hippocampal indexing.[20] Researchers can separately change model weights, temporary activations, context, or external memory, then repeat the same experiment and measure both the intended effect and any side effects. This level of control and repeatability is much harder to achieve in biological systems.

A strong causal effect does not necessarily mean that a fact is stored in one model component; it may instead reveal an important step in a distributed process. Overall, the field has shifted from observing memory-related signals to changing them and measuring intended and unintended effects. Researchers can now ask whether an intervention is selective, lasting, generalizable, reversible, and sensitive to where and when it is applied. **Figure 2** summarizes this progression from early memory theories and neural recordings to selective biological interventions and direct manipulation in LLMs.

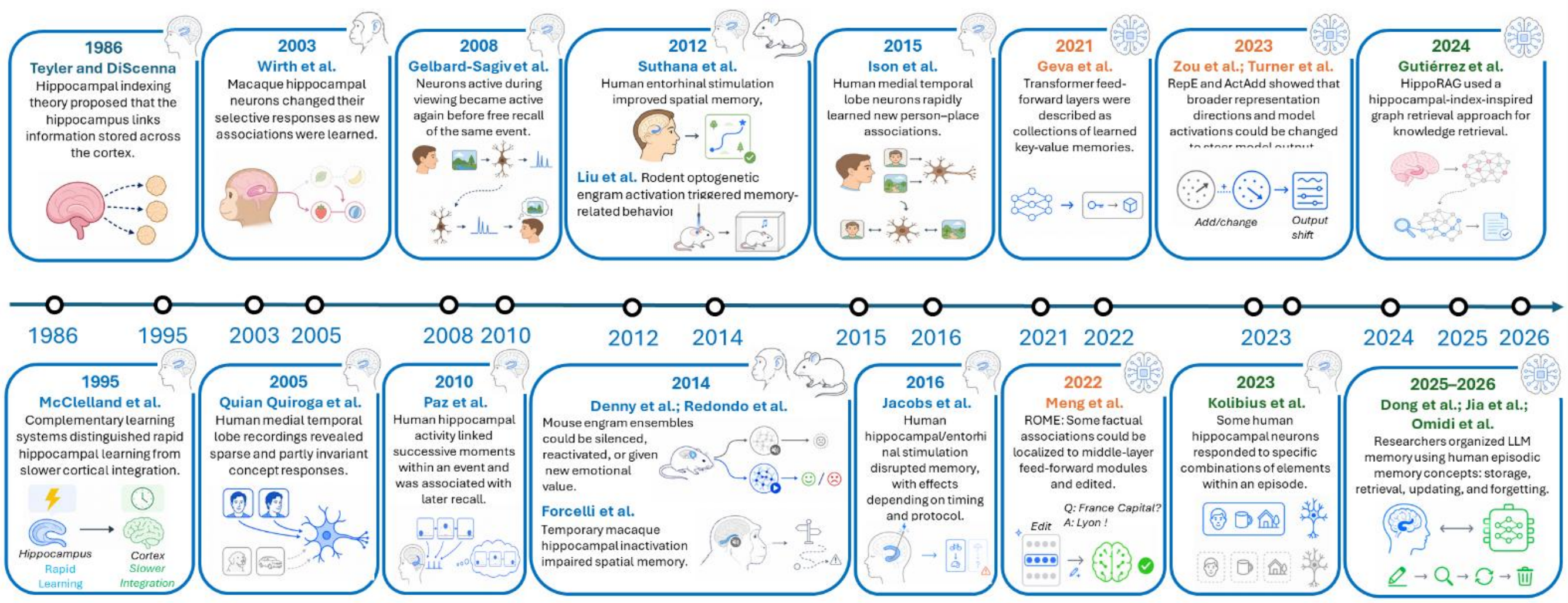

**Figure 2**: Chronological map of evidence across biological and artificial memory systems. Early work proposed hippocampal indexing and complementary learning systems; later studies identified single-neuron coding, temporal binding, engram manipulation, and stimulation effects. More recent work extended similar questions to transformers, model editing, representation steering, and hippocampal-index-inspired retrieval.

## 3. Four shared questions, different mechanisms

The comparison is organized around four questions. Each examines how memory-related information can be identified, measured, and experimentally manipulated across the two systems.

### Representation

Biological memory is distributed across levels: synaptic changes, neuronal ensembles, hippocampal-cortical interactions, and population-level patterns all contribute.[22] Sparse concept-selective or episode-specific neurons are informative, but neither a concept nor an episode should be reduced to one cell.[1,3,5,18] In LLMs, information may be encoded in model weights, expressed through feed-forward and attention computations, represented in residual-stream activations, maintained in the context window, or stored in external memory systems. Feed-forward layers can exhibit key-value-like structure, but their outputs are combined and refined across the network rather than functioning as isolated memory slots.[12,17,23,24] Both fields should study memory at several levels instead of looking for one storage location. A region, neuron, layer, or feature may help retrieve a memory without storing the whole memory.

### Retrieval from partial cues

A partial cue can trigger the brain to recover a more complete memory through interactions between the hippocampus and cortex. Memory-related activity may also appear before a person verbally reports what they remember.[2,4,18] In an LLM, a prompt shapes the model's processing, while LLM systems connected to external documents or databases can provide relevant information without changing the model's weights.[14,20] HippoRAG is one example: inspired by hippocampal indexing, it uses an associative graph to connect and retrieve information across multiple documents.[20]

The shared principle is pattern completion: using incomplete input to recover a fuller representation, although human recall reconstructs a previously experienced event, whereas an LLM may produce an answer from learned statistical associations without retrieving a personal episode.[17]

### Writing and updating

Human medial temporal-lobe neurons can expand their selectivity when a new person-place association is learned, sometimes after very few exposures.[5] Complementary learning systems distinguish rapid learning of new events from their slower integration into stable, generalized knowledge.[19] During ordinary LLM use, model weights are not normally changed. New information may remain in the context window, be added to an external store, or require training or model editing to produce a lasting internal change.[14,15,24]

In both fields, the challenge is not simply to add information. New learning must preserve useful older knowledge and avoid unwanted interference. This balance between learning new information and retaining old information is known as the stability–plasticity problem. The exchange works in both directions. Neuroscience shows that artificial memory systems should consider consolidation, interference, forgetting, and context-dependent updating. LLMs, in turn, provide controlled systems in which researchers can identify and directly test candidate memory representations.

**Intervention**

To compare interventions clearly, both their target and the duration of their effects must be specified. Biological interventions range from selective engram activation or silencing in mice to reversible hippocampal inactivation in macaques and electrical stimulation in humans.[6-11] The same type of intervention can enhance or impair performance depending on the target, timing, frequency, and behavioral state.[10,11] LLM interventions also differ in what they change. ROME produces a persistent weight change to a factual association; ActAdd changes activations during inference; representation engineering shifts broader directions linked to concepts or output tendencies; and RAG changes a separable external store.[13-16] These should not all be described simply as "editing." In both fields, temporary control should be distinguished from lasting change.

## 4. LLMs are ahead in intervention access, not in memory itself

**Table 1** summarizes the current asymmetry in experimental access, including what can be targeted and whether the resulting effects are transient or persistent. Human and macaque studies are grouped because both generally permit less selective intervention than rodent engram studies, despite important differences in methods, behavior, and translational relevance. Rodent studies provide the strongest biological selectivity and uniquely demonstrate causal links between tagged learning-related ensembles and behavior. Human and macaque studies allow more complex behavioral paradigms and closer investigation of human cognition, but their interventions are generally broader. LLMs permit direct, repeatable manipulation of specified artificial components. None of the systems can yet delete one complete memory without changing related knowledge, guarantee an edit with zero unintended effects, or trace the complete causal path from one representation to complex behavior. The LLM advantage described here is therefore experimental access: researchers can manipulate specified components and rerun the system under controlled conditions.

**Table 1:** Intervention capabilities and open challenges across biological and artificial memory systems

| Intervention capability | Human and macaque studies | Rodent studies | LLMs |
| --- | --- | --- | --- |
| Broadly perturb a memory-related system and change behavior or output | ✓ ↺ | ✓ ↺/◆ | ✓ ↺/◆ |
| Change the valence or evaluative tendency of an association | △ ↺/◆ | ✓ ◆ | △ ↺ |
| Selectively manipulate an identified memory-related state | ✗ | ✓ ↺ | ✓ ↺ |
| Make a lasting, targeted change to one association | ✗ | △ ◆ | ✓ ◆ |
| Modify a machine-readable external store automatically accessed during inference, without changing the core system | N/A | N/A | ✓ ◆* |
| Strengthen one association while preserving related ones | ✗ | ✗ | △ ↺/◆ |
| Delete one target memory or association without changing related knowledge | ✗ | ✗ | ✗ |
| Edit one target memory or association with guaranteed zero unintended effects | ✗ | ✗ | ✗ |
| Trace the complete causal path from one representation to complex behavior or output | ✗ | ✗ | ✗ |

Symbols: ✓, capability demonstrated in at least one setting, but not necessarily reliable, perfectly localized, or free of unintended effects; △, partial or strongly context-dependent capability; ✗, not currently demonstrated with high precision; N/A, not meaningfully applicable. Effect duration: ↺, transient or reversible; ◆, persistent after the intervention ends; ↺/◆, both transient and persistent forms have been demonstrated. *For external stores, "persistent" means that the change remains until the stored information is revised or deleted; it does not mean that the model's internal weights have changed.

## 5. Perspective: reverse the experimental pipeline

The open challenges in **Table 1** suggest a different use for LLMs. The hardest goals remain unsolved in every system: deleting one memory without affecting related knowledge, guaranteeing zero unintended effects, and tracing the full causal path from representation to behavior. LLMs may provide an experimentally tractable setting for studying some of these problems because their internal states are easier to measure, perturb, and edit. The goal is to use that control to design sharper biological experiments.

**Questions for the reverse pipeline**

Within this framework, four questions define a practical research agenda:

- How can memory retrieval be distinguished from later response selection, confidence, and report?
- Can one learned association be updated without disrupting related memories?
- What determines whether an intervention produces a temporary change in recall or a lasting memory update?
- Which principles identified in LLMs also appear in mice, macaques, and humans?

This sequence begins by separating retrieval from response, then examines the selectivity and durability of interventions, and ultimately asks which findings generalize across artificial and biological systems.

The familiar direction of influence has often been from biology to AI: human observations identify a phenomenon, animal studies test it under greater control, and computational systems implement a simplified principle. The questions above motivate a complementary direction: LLM intervention → testing in the most suitable biological model → validation in a system with richer behavior → human investigation. The sequence can be adapted depending on the question, required precision, and ethical constraints. The aim is to transfer experimental logic rather than map model parts onto brain parts. This includes what to measure, when to intervene, how long an effect lasts, and what unintended effects appear. **Figure 3** illustrates this proposed reverse experimental pipeline.

As interpretability improves, LLM studies may reveal previously unknown processing stages, representations, or failure modes. These discoveries could generate biological hypotheses that are not obvious from observation alone, while unresolved problems may motivate new methods in both fields. The examples below are framed at the human level, but the same logic could first be tested in mice or macaques when greater intervention selectivity is required.

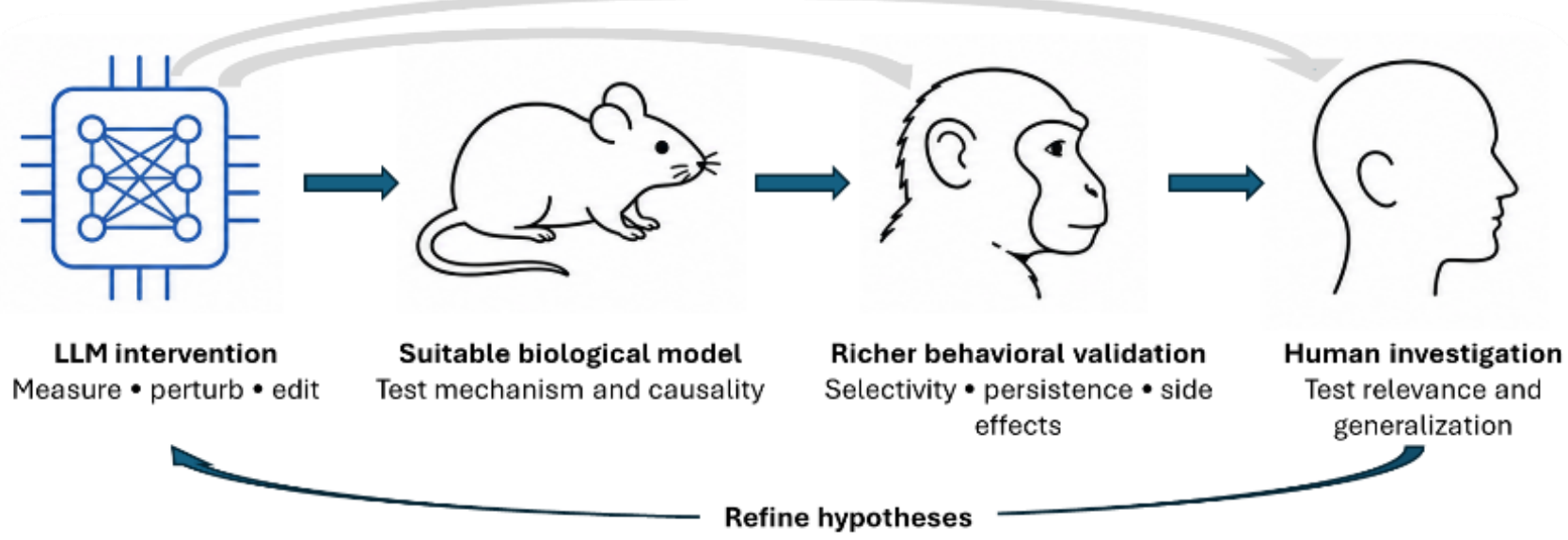

**Figure 3**: The reverse experimental pipeline. LLM interventions generate hypotheses for testing across suitable biological models and, ultimately, humans. The pathway can be adapted to the question, precision required, and ethical constraints.

**Example: tracing competition between old and updated associations**

Meng et al. used causal tracing to identify an early causal site in middle-layer feed-forward modules while the model processed the subject and a later site involving attention near the final output.[15] Human recordings by Gelbard-Sagiv et al. showed that hippocampal and entorhinal neurons could reactivate before a memory was verbally reported.[2] However, the complete sequence from cue to final answer remains unclear in both LLMs and brains.

One caution is that layer order and token order represent different kinds of timing in an LLM. For each new token, the model processes the available text through the same sequence of layers, from early to late. It then generates one token, adds it to the context, and repeats the process for the next token. Thus, early and late layers describe computational steps within one prediction, while earlier and later tokens describe how the text and context develop over time. Neither should be treated as a direct match to a specific brain region or biological time point. The following are possible future LLM findings and the biological hypotheses they could motivate:

- If an outdated association appears early and remains detectable even after the updated answer becomes dominant, researchers could test whether the brain first reactivates an older memory and whether that memory remains represented after it no longer controls the response.
- If different model components contribute to different stages of updating, and interventions at those stages produce different effects, researchers could test whether different neural populations contribute to retrieval and later response selection, and whether intervention during an early retrieval-related stage has different effects from intervention during selection, confidence, or report.
- If a later or incomplete shift from the old association to the updated one predicts incorrect or unstable outputs, and this pattern differs across LLMs, researchers could test whether people likewise vary in how quickly an updated memory becomes dominant and whether these differences predict intrusion errors, delayed responses, or lower confidence.

This approach moves the question from whether a memory was reactivated to how old and updated associations emerge, compete, and influence the final response.

### A shared evaluation language for memory updating

LLM editing research evaluates more than whether the target answer changed; it also asks whether the change affects only the intended association, works across reworded prompts, preserves unrelated knowledge, lasts over time, and avoids unintended effects. More broadly, memory should be evaluated not only by whether recall is correct, but also by how well information is retrieved, updated, retained, forgotten, and changed by an intervention. Biological memory-updating studies could use the same evaluation criteria without assuming that the underlying mechanisms are the same. For example, after changing a learned context → threat association into context → safety, researchers could test whether the update generalizes to similar cues, leaves unrelated threat memories unchanged, affects connected memories, persists over time, returns under stress or context change, or alters overlapping neuronal ensembles. The framework should also separate temporary changes that last only during the intervention from lasting changes that remain afterward. Together, these criteria could form a shared benchmark for LLM and biological memory interventions. It would ask whether an intervention works, affects only the intended association, generalizes to related cues, lasts or can be reversed, causes unintended changes, and whether its effects can be traced to internal changes. The benchmark should report these results separately, using measures appropriate to each system.

### Guardrails for the comparison

The proposed reverse pipeline depends on four guardrails. First, an LLM result is a hypothesis generator, not evidence about the brain. Second, the comparison should focus on shared functions and experimental questions, not on matching brain regions to model components. Third, outputs alone are not enough: interventions should also be tested for unintended effects, changes to related knowledge, and how long the effect lasts. Fourth, null results are informative. A principle that works in an LLM may fail in mice, macaques, or humans because biological memory is shaped by the body, emotion, neuromodulation, development, and lived experience.

These guardrails keep the comparison focused on testable functional questions. They support productive translation between LLM and biological memory research while respecting the distinct mechanisms of each.

## Conclusion

Brains and LLMs are most usefully compared through shared functional questions, not assumed biological equivalence. Neuroscience shows how memory is grounded in experience, emotion, perception, and behavior, whereas LLM research provides unusually direct access to internal representations and allows them to be measured, perturbed, and edited. LLMs may therefore offer an experimentally tractable setting for studying problems that remain difficult in biological systems, including selective updating, persistence, reversibility, and unintended effects. This access can help generate sharper biological hypotheses. After decades in which neuroscience inspired AI, the direction can now partly reverse: discoveries in LLMs can motivate targeted experiments in mice, macaques, and humans. The next frontier is not a better metaphor between brains and models, but a stronger experimental cycle: identify a principle in an LLM, intervene on it, measure the result, and test whether a related process appears in biological memory. Future LLM discoveries may not only refine existing experiments but also reveal new questions about biological memory itself.

**Acknowledgments**

This research was partially supported by CPRIT RP240117. The funding sources had no role in the study design; data collection, analysis, or interpretation; or manuscript preparation.

**Author contributions**

M.S. conceived the original idea. A.M. and S.A.Z. contributed to the development and refinement of the concept. M.S., S.A.Z., A.M. and J.W. contributed to the literature review, writing, visualization and overall development of the manuscript. J.W. supervised the work and provided critical review and editing. All authors reviewed and approved the final version of the manuscript.

**Competing interests:** The authors declare no competing interests.

**Additional information**: Correspondence should be addressed to Jia Wu.